\documentclass[12pt]{article}
\usepackage{amssymb}
\usepackage{amsfonts}
\usepackage{amsmath}
\usepackage{epsfig}

\begin{document}

\title{The fractional volatility model: No-arbitrage, leverage and risk
measures}
\author{R.~Vilela Mendes\thanks{%
Centro de Matem\'{a}tica e Aplica\c{c}\~{o}es Fundamentais, Universidade de
Lisboa, Av. Prof. Gama Pinto 2, P1649-003 Lisboa} \thanks{%
IPFN, EURATOM/IST Association, e-mail: vilela@cii.fc.ul.pt}\quad and Maria Jo%
\~{a}o Oliveira\footnotemark[1] \thanks{%
Universidade Aberta, oliveira@cii.fc.ul.pt}}
\date{}
\maketitle

\begin{abstract}
Based on a criterium of mathematical simplicity and consistency with
empirical market data, a stochastic volatility model has been obtained with
the volatility process driven by fractional noise. Depending on whether the
stochasticity generators of log-price and volatility are independent or are
the same, two versions of the model are obtained with different leverage
behavior. Here, the no-arbitrage and incompleteness properties of the model
are studied. Some risk measures are also discussed in this framework.
\end{abstract}


\section{Introduction}

In liquid markets the autocorrelation of price changes decays to negligible
values in a few minutes, consistent with the absence of long term
statistical arbitrage. Because innovations of a martingale are uncorrelated,
there is a strong suggestion that it is a process of this type that should
be used to model the stochastic part of the returns process. Classical
Mathematical Finance has, for a long time, been based on the assumption that
the price process of market securities may be approximated by geometric
Brownian motion 
\begin{equation}
\begin{array}{lll}
dS_{t} & = & \mu S_{t}dt+\sigma S_{t}dB\left( t\right)%
\end{array}
\label{1.1}
\end{equation}%
Geometric Brownian motion (GBM) models the absence of linear correlations,
but otherwise has some serious shortcomings. It does not reproduce the
empirical leptokurtosis nor does it explain why nonlinear functions of the
returns exhibit significant positive autocorrelation. For example, there is
volatility clustering, with large returns expected to be followed by large
returns and small returns by small returns (of either sign). This, together
with the fact that autocorrelations of volatility measures decline very
slowly \cite{Ding2}, \cite{Harvey}, \cite{Crato} has the clear implication
that long memory effects should somehow be represented in the process and
this is not included in the geometric Brownian motion hypothesis. The
existence of an essential memory component is also clear from the failure of
reconstruction of a Gibbs measure and the need to use chains with complete
connections in the phenomenological reconstruction of the market process 
\cite{Vilela1}.

As pointed out by Engle \cite{Engle}, when the future is uncertain investors
are less likely to invest. Therefore uncertainty (volatility) would have to
be changing over time. The conclusion is that a dynamical model for
volatility is needed and $\sigma $ in Eq.(\ref{1.1}), rather than being a
constant, becomes itself a process. This idea led to many deterministic and
stochastic models for the volatility (\cite{Taylor}, \cite{Engle2} and
references therein).

The stochastic volatility models that were proposed described some partial
features of the market data. For example leptokurtosis is easy to fit but
the long memory effects are much harder. On the other hand, and in contrast
with GBM, some of the phenomenological fittings of historical volatility
lack the kind of nice mathematical properties needed to develop the tools of
mathematical finance. In an attempt to obtain a model that is both
consistent with the data and mathematically sound, a new approach was
developed in \cite{Oliveira}. Starting only with some criteria of
mathematical simplicity, the basic idea was to let the data itself tell us
what the processes should be.

The basic hypothesis for the model construction were:

(H1) The log-price process $\log S_{t}$ belongs to a probability product
space $(\Omega _{1}\times \Omega _{2},P_{1}\times P_{2})$ of which the $%
(\Omega _{1},P_{1})$ is the Wiener space and the second one, $(\Omega
_{2},P_{2})$, is a probability space to be reconstructed from the data.
Denote by $\omega _{1}\in \Omega _{1}$ and $\omega _{2}\in \Omega _{2}$ the
elements (sample paths) in $\Omega _{1}$ and $\Omega _{2}$ and by $\mathcal{F%
}_{1,t}$ and $\mathcal{F}_{2,t}$ the $\sigma $-algebras in $\Omega _{1}$ and 
$\Omega _{2}$ generated by the processes up to $t$. Then, a particular
realization of the log-price process is denoted 
\begin{equation*}
\log S_{t}\left( \omega _{1},\omega _{2}\right)
\end{equation*}%
This first hypothesis is really not limitative. Even if none of the
non-trivial stochastic features of the log-price were to be captured by
Brownian motion, that would simply mean that $S_{t}$ was a trivial function
in $\Omega _{1}$.

(H2) The second hypothesis is stronger, although natural. It is assumed that
for each fixed $\omega_2$, $\log S_{t}\left( \cdot ,\omega_2\right) $ is a
square integrable random variable in $\Omega_1$.

These principles and a careful analysis of the market data led, in an
essentially unique way\footnote{%
Essentially unique in the sense that the empiricaly reconstructed volatility
process is the simplest one, consistent with the scaling properties of the
data.}, to the following model:

\begin{equation}
\begin{array}{lll}
dS_{t} & = & \mu S_{t}\,dt+\sigma _{t}S_{t}\,dB\left( t\right) \\ 
\log \sigma _{t} & = & \beta +\frac{k}{\delta }\left\{ B_{H}\left( t\right)
-B_{H}\left( t-\delta \right) \right\}%
\end{array}
\label{1.2}
\end{equation}%
the data suggesting values of $H$ in the range $0.8-0.9$. In this coupled
stochastic system, in addition to a mean value, volatility is driven by
fractional noise. Notice that this empirically based model is different from
the usual stochastic volatility models which assume the volatility to follow
an arithmetic or geometric Brownian process. Also in Comte and Renault \cite%
{Comte} and Hu \cite{Hu}, it is fractional Brownian motion that drives the
volatility, not its derivative (fractional noise). $\delta $ is the
observation scale of the process. In the $\delta \rightarrow 0$ limit the
driving process would be the distribution-valued process $W_{H}$%
\begin{equation}
W_{H}=\lim_{\delta \rightarrow 0}\frac{1}{\delta }\left( B_{H}\left(
t\right) -B_{H}\left( t-\delta \right) \right)  \label{1.3}
\end{equation}%
The second equation in (\ref{1.2}) leads to 
\begin{equation}
\sigma \left( t\right) =\theta e^{\frac{k}{\delta }\left\{ B_{H}\left(
t\right) -B_{H}\left( t-\delta \right) \right\} -\frac{1}{2}\left( \frac{k}{%
\delta }\right) ^{2}\delta ^{2H}}  \label{1.4}
\end{equation}%
with $E\left[ \sigma \left( t\right) \right] =\theta >0$.

The model has been shown \cite{Oliveira} to describe well the statistics of
price returns for a large $\delta$-range and a new option pricing formula,
with "smile" deviations from Black-Scholes, was also obtained. Here we will
be concerned with general consistency questions for the model, namely
arbitrage and market completeness. Also, in Section 3, some new results on
risk measures will be presented.

\section{No-arbitrage and market incompleteness}

It had been clear for a long time that the slow decline of the volatility
autocorrelations implied the existence of some kind of long memory effect in
the market. Several authors tried to describe this effect by replacing in
the price process Brownian motion by fractional Brownian motion with $H>1/2$%
. However it was soon realized \cite{Rogers}, \cite{Shiryaev}, \cite{Salopek}%
, \cite{Sottinen} that this replacement implied the existence of arbitrage.
These results might be avoided either by restricting the class of trading
strategies \cite{Cheridito}, introducing transaction costs \cite{Guasoni} or
replacing pathwise integration by a different type of integration \cite%
{Oksendal} \cite{Elliot}. However this is not free of problems because the
Skorohod integral approach requires the use of a Wick product either on the
portfolio or on the self-financing condition, leading to unreasonable
situations from the economic point of view (for example positive portfolio
with negative Wick value, etc.) \cite{Hult}.

The fractional volatility model in (\ref{1.2}) is not affected by these
considerations, because it is the volatility process that is driven by
fractional noise, not the price process. In fact a no-arbitrage result may
be proven. This is no surprise because our requirement (H2) that, for each
sample path $\omega _{2}\in \Omega _{2}$, $\log S_{t}\left( \cdot ,\omega
_{2}\right) $ is a square integrable random variable in $\Omega _{1}$
already implies that $\int \sigma _{t}dB_{t}$ is a martingale. The square
integrability is also essential to guarantee the possibility of
reconstruction of the $\sigma $ process from the data, because it implies 
\cite{Nualart}%
\begin{equation}
\begin{array}{lll}
\frac{dS_{t}}{S_{t}}\left( \cdot ,\omega _{2}\right) & = & \mu _{t}\left(
\cdot ,\omega _{2}\right) dt+\sigma _{t}\left( \cdot ,\omega _{2}\right)
dB_{t}%
\end{array}
\label{2.2}
\end{equation}%
We now consider a market with an asset obeying the stochastic equations (\ref%
{1.2}) and a risk-free asset $A_{t}$%
\begin{equation}
dA_{t}=rA_{t}\,dt  \label{2.3}
\end{equation}%
with $r>0$ constant.

\noindent \textbf{Proposition 1:} \textit{The market defined by (\ref{1.2})
and (\ref{2.3}) is free of arbitrages}

The proofs of this and the next proposition follow the same steps as in \cite%
{Kallianpur}. Technically, the similarity of the proofs follows from the
properties of volatility process (\ref{1.4}).

\smallskip

\noindent \textbf{Lemma:} \textit{For }$\sigma $\textit{\ given by (\ref{1.4}%
) one has:\newline
i) For any integer number }$n$\textit{, }$\int_{0}^{T}\mathbb{E}\left(
\sigma _{t}^{n}\right) dt<\infty $, \textit{where the expectation is with
respect to the probability measure }$P_{2}$\textit{;\newline
ii) Assuming that }$\mu \in L^{\infty }\left( \left[ 0,T\right] ,P_{1}\times
P_{2}\right) $\textit{\footnote{%
Since this assumption is quite natural, one assumes it throughout this work.
In addition, we also assume that $\mu _{t}$ is adapted to the product
filtration $\mathcal{F}_{1,t}\times \mathcal{F}_{2,t}$.}, for any }$t\in %
\left[ 0,T\right] $\textit{\ there is a constant }$C>0$\textit{\ such that }$%
P_{1}\times P_{2}$\textit{-a.e.} 
\begin{equation*}
\int_{0}^{t}\frac{(r-\mu _{s})^{2}}{\sigma _{s}^{2}}\,ds\leq C
\end{equation*}

\smallskip

\noindent \textbf{Proof:} The first property follows from 
\begin{equation*}
\mathbb{E}\left(e^{\lambda\left( B_{H}(t) -B_{H}( t-\delta)\right)}\right)
=e^{\frac{\lambda^2}{2}\delta^{2H}}
\end{equation*}
for any complex number $\lambda$, while the second one from the H\"older
continuity of the fractional Brownian motion $B_H$ of order less than $H$
(cf.~\cite{Decreusefond}). More precisely, for each $\alpha\in\left(0,H%
\right)$ there is a constant $C_\alpha>0$ such that $P_2$-a.e. 
\begin{equation*}
\left|B_{H}(t) -B_{H}(s)\right|\leq C_\alpha\left| t-s\right|^\alpha
\end{equation*}
and thus $P_1\times P_2$-a.e. 
\begin{eqnarray*}
\int_0^t\frac{(r-\mu_s)^2}{\sigma_s^2}\,ds&\leq& \frac{(r+\Vert
\mu\Vert_\infty)^2}{\theta^2}\,e^{\left( \frac{k}{\delta }\right) ^{2}\delta
^{2H}}\int_0^t e^{\frac{2k}{\delta }\left|B_{H}(s) -B_{H}(
s-\delta)\right|}\,ds \\
&\leq& \frac{T(r+\Vert \mu\Vert_\infty)^2}{\theta^2}\, e^{\left( \frac{k}{%
\delta }\right) ^{2}\delta ^{2H} +2kC_\alpha\delta^{\alpha-1}}
\end{eqnarray*}
\hfill$\blacksquare \medskip$

\noindent \textbf{Proof of Proposition 1: } Let $P:=P_{1}\times P_{2}$ be
the probability product measure and define the process%
\begin{equation}
Z_{t}=\frac{S_{t}}{A_{t}}  \label{2.4}
\end{equation}%
in the interval $0\leq t\leq T$, which obeys the equation%
\begin{equation}
dZ_{t}=\left( \mu _{t}-r\right) Z_{t}\,dt+\sigma _{t}Z_{t}\,dB_{t}
\label{2.5}
\end{equation}%
Now let%
\begin{equation}
\eta _{t}=\exp \left( \int_{0}^{t}\frac{r-\mu _{s}}{\sigma _{s}}\,dB_{s}-%
\frac{1}{2}\int_{0}^{t}\frac{\left\vert r-\mu _{s}\right\vert ^{2}}{\sigma
_{s}^{2}}\,ds\right)  \label{2.6}
\end{equation}%
which by the Lemma fulfills the Novikov condition and thus it is a $P$%
-martingale. Moreover, it yields a probability measure $P^{\prime }$
equivalent to $P$ by 
\begin{equation}
\frac{dP^{^{\prime }}}{dP}=\eta _{T}  \label{2.7}
\end{equation}%
By the Girsanov theorem 
\begin{equation}
B_{t}^{\ast }=B_{t}-\int_{0}^{t}\frac{r-\mu _{s}}{\sigma _{s}}\,ds
\label{2.8}
\end{equation}%
is a $P^{\prime }-$Brownian motion and%
\begin{equation}
Z_{t}=Z_{0}+\int_{0}^{t}\sigma _{s}Z_{s}\,dB_{s}^{\ast }  \label{2.9a}
\end{equation}%
is a $P^{\prime }$-martingale. By the fundamental theorem of asset pricing,
the existence of an equivalent martingale measure for $Z_{t}$ implies that
there are no arbitrages, that is, $\mathbb{E}_{P^{\prime }}\left[ Z_{t}|%
\mathcal{F}_{1,s}\times \mathcal{F}_{2,s}\right] =Z_{s}$ for $0\leq s<t\leq
T $.\hfill $\blacksquare \medskip $

\noindent \textbf{Proposition 2:} \textit{The market defined by (\ref{1.2})
and (\ref{2.3}) is incomplete}

\smallskip

\noindent \textbf{Proof: }Here we use an integral representation for the
fractional Brownian motion \cite{Decreusefond}, \cite{Embrechts} 
\begin{equation}
B_{H}\left( t\right) =C\int_{0}^{t}K\left( t,s\right) dW_{s}  \label{2.10a}
\end{equation}%
$W_{t}$ being a Brownian motion independent from $B_{t}$ and $K$ the square
integrable kernel 
\begin{equation*}
K\left( t,s\right) =C_{H}s^{\frac{1}{2}-H}\int_{s}^{t}(u-s)^{H-\frac{3}{2}%
}u^{H-\frac{1}{2}}\,du,\quad s<t
\end{equation*}%
($H>1/2)$. Let $(B_{t},W_{t})$ be a bi-dimensional Brownian motion on $P$.
Given the $P_{2}$-martingale 
\begin{equation}
\eta _{t}^{\prime }=\exp \left( W_{t}-\frac{1}{2}t\right)  \label{2.11a}
\end{equation}%
we now use the product $\eta _{t}\eta _{t}^{\prime }$. Due to the Lemma, the
Novikov condition is fulfilled insuring that $\eta _{t}\eta _{t}^{\prime }$
is a $P$-martingale and 
\begin{equation}
\frac{dP^{\prime \prime }}{dP}=\eta _{T}\eta _{T}^{\prime }  \label{2.12a}
\end{equation}%
a probability measure. As before, the Girsanov theorem implies that the $%
Z_{t}$ process is still a $P^{\prime \prime }$-martingale. The equivalent
martingale measure not being unique the market is, by definition,
incomplete.\hfill $\blacksquare \medskip $

Incompleteness of the market is a reflection of the fact that in stochastic
volatility models there are two different sources of risk and only one of
the risky assets is traded. In this case a choice of measure is how one
fixes the volatility risk premium.

\section{Leverage and the identification of the stochastic generators}

The following nonlinear correlation of the returns 
\begin{equation}
L\left( \tau \right) =\left\langle \left\vert r\left( t+\tau \right)
\right\vert ^{2}r\left( t\right) \right\rangle -\left\langle \left\vert
r\left( t+\tau \right) \right\vert ^{2}\right\rangle \left\langle r\left(
t\right) \right\rangle  \label{C1}
\end{equation}%
is called \textit{leverage} and the \textit{leverage effect} is the fact
that, for $\tau >0$, $L\left( \tau \right) $ starts from a negative value
whose modulus decays to zero whereas for $\tau <0$ it has almost negligible
values. In the form of Eqs.~(\ref{1.2}) the volatility process $\sigma _{t}$
affects the log-price, but is not affected by it. Therefore, in its simplest
form the fractional volatility model contains no leverage effect.

Leverage may, however, be implemented in the model in a simple way \cite%
{Vilela3} if one identifies the Brownian processes $B_{t}$ and $W_{t}$ in (%
\ref{1.2}) and (\ref{2.10a}). Identifying the random generator of the
log-price process with the stochastic integrator of the volatility, at least
a part of the leverage effect is taken into account.

The identification of the two Brownian processes means that now, instead of
two, there is only one source of risk. Hence it is probable that in this
case completeness of the market might be achieved. However questions like
mathematical consistency and arbitrage properties of the new model are to be
checked.

Let us now consider the market (\ref{1.2}) and (\ref{2.3}) with $B_{t}$
appearing in (\ref{1.2}) replaced by the standard Brownian motion $W_{t}$
which appears in the integral representation (\ref{2.10a}).

\noindent \textbf{Proposition 3:} \textit{This new market is free of
arbitrages}

\smallskip

\noindent \textbf{Proof:} In this case $P_{1}=P_{2}$. Since the item ii) in
the Lemma still holds for the product measure $P_{1}\times P_{2}$ replaced
by the probability measure $P_{2}$, with this change of probability measure
the proof of this result follows as in the proof of Proposition 1.\hfill $%
\blacksquare \medskip $

\section{Some risk measures}

Let $\delta S=S_{t+\Delta }-S_{t}$ and%
\begin{equation}
r\left( \Delta \right) =\log S_{t+\Delta }-\log S_{t}  \label{3.0}
\end{equation}%
be the return corresponding to a time lag $\Delta $.

The value at risk (VaR) $\Lambda ^{\ast }$ and the expected shortfall $%
E^{\ast }$ are%
\begin{equation}
\int_{-S}^{-\Lambda ^{\ast }}P_{\Delta }\left( \delta S\right) d\left(
\delta S\right) =P^{\ast }  \label{3.4}
\end{equation}%
\begin{equation}
E^{\ast }=\frac{1}{P^{\ast }}\int_{-S}^{-\Lambda ^{\ast }}\left( -\delta
S\right) P_{\Delta }\left( \delta S\right) d\left( \delta S\right)
\label{3.5}
\end{equation}%
$S$ being the capital at time zero, $P^{\ast }$ the probability of a loss $%
\Lambda ^{\ast }$ and $P_{\Delta }\left( \delta S\right) $ the probability
of a price variation $\delta S$ in the time interval $\Delta $. In terms of
the returns these quantities are%
\begin{equation}
\int_{-\infty }^{\log \left( 1-\frac{\Lambda ^{\ast }}{S}\right) }P\left(
r\left( \Delta \right) \right) d\left( r\left( \Delta \right) \right)
=P^{\ast }  \label{3.6}
\end{equation}%
\begin{equation}
E^{\ast }=\frac{S}{P^{\ast }}\int_{-\infty }^{\log \left( 1-\frac{\Lambda
^{\ast }}{S}\right) }\left( 1-e^{r\left( \Delta \right) }\right) P\left(
r\left( \Delta \right) \right) d\left( r\left( \Delta \right) \right)
\label{3.7}
\end{equation}%
For the fractional volatility model the probability distribution of the
returns in a time interval $\Delta $, is obtained \cite{Oliveira} from%
\begin{equation}
P\left( r\left( \Delta \right) \right) =\int_{0}^{\infty }p_{\delta }\left(
\sigma \right) p_{\sigma }\left( r\left( \Delta \right) \right) d\sigma
\label{3.3}
\end{equation}%
with 
\begin{equation}
p_{\delta }\left( \sigma \right) =\frac{1}{\sqrt{2\pi }\sigma k\delta ^{H-1}}%
\exp \left\{ -\frac{\left( \log \sigma -\beta \right) ^{2}}{2k^{2}\delta
^{2H-2}}\right\}  \label{3.1}
\end{equation}%
\begin{equation}
p_{\sigma }\left( r\left( \Delta \right) \right) =\frac{1}{\sqrt{2\pi \sigma
^{2}\Delta }}\exp \left\{ -\frac{\left( r\left( \Delta \right) -\left( \mu -%
\frac{\sigma ^{2}}{2}\right) \Delta \right) ^{2}}{2\sigma ^{2}\Delta }%
\right\}  \label{3.2}
\end{equation}

Using (\ref{3.3})--(\ref{3.2}) $\Lambda ^{\ast }$ and $E^{\ast }$ are
computed from (\ref{3.6}) and (\ref{3.7}). As an illustration in the figures
1 and 2 one shows the results for $P^{\ast }=0.01$ ($99\%$VaR) and time lags
from 1 to 30 days, using the following parameters%
\begin{equation*}
H=0.83;k=0.59,\beta =-5,\delta =1
\end{equation*}%
These values are obtained from typical market daily data ($\delta =1$ day).
The results are also compared with those obtained from a simple lognormal
price distribution with the same averaged volatility.

\begin{figure}[tbh]
\begin{center}
\psfig{figure=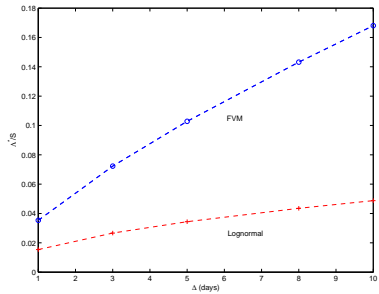,width=5truecm}
\end{center}
\caption{VaR in the fractional volatility model compared with the lognormal 
with the same average volatility}
\end{figure}

\begin{figure}[tbh]
\begin{center}
\psfig{figure=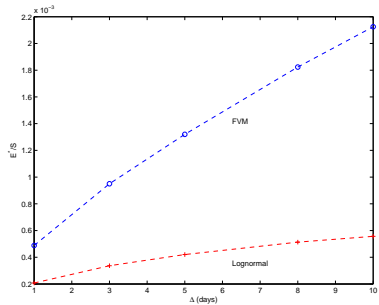,width=5truecm}
\end{center}
\caption{Expected shortfall in the fractional volatility model compared 
with the lognormal with the same average volatility}
\end{figure}

One sees that both for VaR and the expected shortfall, the fractional
volatility model predicts much larger values than the lognormal. This
results mostly from the fatter tails in the model (as well as in the market
data).

\section{Remarks and conclusions}

1) Being partially reconstructed from empirical data, it is no surprise that
the fractional volatility model describes well the statistics of returns.
The fact that, once the parameters are adjusted by the data for a particular
observation time scale $\delta $, it describes well very different time lags
seems to be related to the fact that the volatility is driven not by
fractional Brownian motion but its increments.

2) What at first seemed surprising was the fact that the same set of
parameters would describe very different markets \cite{Oliveira}. This
motivated a search for the kind of behavior of the market agents which would
be consistent with the statistical properties observed in the model (and
also on the empirical data). Two stylized models were considered. In the
first the traders strategies play a determinant role. In the second the
determinant effect is the limit-order book dynamics, the agents having a
random nature. The conclusion was that the market statistical behavior (in
normal days) seems to be more influenced by the nature of the financial
institutions (the double auction process) than by the traders strategies 
\cite{Vilela2}. Specific trader strategies and psychology should however
play a role on market crisis and bubbles. Therefore some kind of
universality of the statistical behavior of the bulk data throughout
different markets would not be surprising.

3) As pointed out in Section 3, the identification of the Brownian process
of the log-price with the one that generates the fractional noise driving
the volatility, introduces an asymmetric coupling between $\sigma _{t}$ and $%
S_{t}$ that is also exhibited by the market data. It is natural to expect
that in this case, because there is only one generator of stochastic risk,
the market will be complete. A rigorous mathematical proof of this result,
which would be akin to the proof of uniqueness of a constrained Girsanov
construction, is still lacking.

\end{document}